\documentclass[aps, pre, twocolumn,showkeys,showpacs,amsmath,amssymb,superscriptaddress]{revtex4-1}

\usepackage{lineno,hyperref}
\modulolinenumbers[5]

\usepackage{todonotes}
\usepackage{makeidx}
\usepackage{graphics}
\usepackage{color}
\usepackage{listings}
\usepackage{bm}
\usepackage{url}
\usepackage{booktabs}
\usepackage{amsmath}
\usepackage{paralist}
\usepackage{amssymb}
\usepackage{amsthm}
\usepackage{subcaption}
\usepackage{algorithm}
\usepackage{algorithmic}

\usepackage[percent]{overpic}

\graphicspath{ {img/} {plot/} }

\newcommand*\diff{\mathop{}\!\mathrm{d}}

\begin{document}


\title{Relativistic dissipation obeys Chapman-Enskog asymptotics:
       Analytical and numerical evidence as a basis for accurate kinetic simulations}
\author{A. Gabbana}
\affiliation{Universit\`a di Ferrara and INFN-Ferrara, I-44122 Ferrara,~Italy}
\affiliation{Bergische Universit\"at Wuppertal, D-42119 Wuppertal,~Germany}
\author{D. Simeoni}
\affiliation{Universit\`a di Ferrara and INFN-Ferrara, I-44122 Ferrara,~Italy}
\affiliation{Bergische Universit\"at Wuppertal, D-42119 Wuppertal,~Germany}
\affiliation{University of Cyprus, CY-1678 Nicosia,~Cyprus}
\author{S. Succi}
\affiliation{Center for Life Nano Science @ La Sapienza, Italian Institute of Technology, Viale Regina Elena 295, I-00161 Roma,~Italy}
\affiliation{Istituto Applicazioni del Calcolo, National Research Council of Italy, Via dei Taurini 19, I-00185 Roma,~Italy}
\author{R. Tripiccione}
\affiliation{Universit\`a di Ferrara and INFN-Ferrara, I-44122 Ferrara,~Italy}

\begin{abstract}
We present an analytical derivation of the transport coefficients of a 
relativistic gas in $(2+1)$ dimensions for both Chapman-Enskog 
(CE) asymptotics and Grad's expansion methods. 
We further develop a systematic calibration method, connecting the relaxation time of relativistic kinetic
theory to the transport parameters of the associated dissipative hydrodynamic equations.
Comparison of our analytical results and numerical simulations shows that 
the CE method correctly captures dissipative effects, while Grad's method does not, in agreement with previous analyses performed in the $(3+1)$ dimensional case.
These results provide a solid basis for accurately calibrated computational 
studies of  relativistic dissipative flows.
\end{abstract}

\maketitle

\section{Introduction}\label{sec:intro}

In the recent years, relativistic fluid dynamics \cite{rezzolla-book-2013} has met with a major surge of interest, due to its crucial  
role in several areas of modern physics, such as the transport properties of high-temperature astrophysical plasmas \cite{uzdensky-rpp-2014}, 
dark-matter cosmology \cite{hui-prd-2017} and the dynamics of quark-gluon plasmas 
in high-energy heavy-ion collisions \cite{baier-prc-2006,romatschke-arxiv-2017}.

In this context, there is major scope for developing efficient and accurate numerical solvers for the study of 
dissipative relativistic hydrodynamics, since controlled experimental setups are often not viable, while 
analytical methods suffer major limitations in describing complex phenomena which arise from 
strong nonlinearities and/or non-ideal geometries of direct relevance for experiments.
In the last decade, mesoscale lattice kinetic schemes \cite{mendoza-prl-2010,romatschke-prc-2011,gabbana-pre-2017} 
have emerged as a promising tool for the study of dissipative hydrodynamics in relativistic regimes. 

One of the assets of the kinetic approach is that the emergence of viscous effects does not 
break relativistic invariance and causality, because space and time are treated on the same footing, i.e. both 
via first order derivatives (hyperbolic formulation).
This overcomes many conceptual issues associated with the consistent formulation
of relativistic transport phenomena. 
Indeed, it is well known that a straightforward relativistic extension of the Navier-Stokes equations is inconsistent with 
relativistic invariance, because second order space derivatives imply superluminal propagation, hence non-causal and unstable behaviour. 
In 1979 Israel and Stewards (IS) introduced a hyperbolic formulation \cite{israel-anp-1976,israel-prsl-1979} able to restore causal dissipation, thus providing a valuable reference framework for subsequent studies to this day. 
However, recent work has highlighted both theoretical shortcomings \cite{denicol-prd-2012} of the IS formulation,  
as well as poor agreement with numerical solutions of the Boltzmann equation 
\cite{huovinen-prc-2009, bouras-prc-2010,florkowski-prc-2013}.
Several alternative formulations have been proposed in recent years \cite{muronga-prc-2007,muronga-prc-2007b,betz-epj-2009,el-prc-2010,
denicol-prl-2010,betz-epj-2011,denicol-prd-2012,jaiswal-prc-2013,jaiswal-prc-2013b,jaiswal-prc-2013c,bhalerao-prc-2013,
bhalerao-prc-2014,chattopadhyay-prc-2015}, but a consistent definition of a causal theory of relativistic viscous
hydrodynamics and the accurate determination of the associated transport coefficients, is still under debate.

The IS formulation follows from the Boltzmann equation, using a relativistic extension of Grad's moments method, 
\cite{grad-cpam-1949}, commonly used to derive hydrodynamic equations from the Boltzmann equation.
Grad's method is not the only route from kinetic theory to hydrodynamics,
another viable alternative being provided by the Chapman-Enskog (CE) expansion \cite{chapman-book-1970}. 

The two differ significantly in spirit and technical detail as well: Grad's method is basically 
an expansion of the Boltzmann probability distribution function in Hilbert space, which is usually
truncated at the level of the third kinetic moment (energy flux).
Chapman-Enskog asymptotics, on the other hand, is a multi-scale expansion based 
on a weak-gradient approximation, i.e. weak departure from local equilibrium.

Both procedures come with well-known limitations: Grad's truncation endangers positive-definiteness of
the distribution function, while the Chapman-Enskog expansion suffers convergence problems
in the presence of strong gradients, or, more precisely, whenever the heterogeneity scale
of hydrodynamic fields becomes comparable with the molecular mean free path (finite Knudsen number).

Despite these differences and limitations, in the non-relativistic regime,
both methods connect kinetic theory and hydrodynamics in a
consistent way,  i.e. they provide the same transport coefficients. 
Yet, in the relativistic regime, this is no longer the case and the immediate question
arises as to which (if any) of the two provides the correct  description of the hydrodynamic limit. 

This question has been studied by several authors, at
the theoretical level \cite{denicol-prl-2010, denicol-prd-2012, molnar-prd-2014, jaiswal-prc-2013b, tsumura-epj-2012, tsumura-prd-2015, kikuchi-prc-2015, kikuchi-pla-2016},
but only very recently has this extensive analysis -- complemented by results of numerical simulations \cite{plumari-prc-2012,florkowski-prc-2013,bhalerao-prc-2014} -- decidedly pointed in favour of the CE procedure; 
All these analyses are restricted to three-dimensional fluids in the ultra-relativistic limit, 
with virtually no results available in the mildly relativistic regime or for the two-dimensional case. A notable exception in $(3+1)$ dimensions \cite{gabbana-pre-2017b} shows that numerical simulations are able to clearly discriminate between CE and Grad's method on a wide range of kinematic regimes and neatly confirms that the CE approach is the correct one.
While the (3+1)-dimensional case is obviously relevant in terms of potential
applications, the study of relativistic fluids in lower dimensions may be of
practical interest since  it is considerably simpler to handle both at a
mathematical \cite{kremer-prd-2002} and computational level
\cite{kellerman-cqg-2008}.

More interestingly, it has been recently realised that two-dimensional
relativistic fluid dynamics   captures several aspects of the collective
dynamics of exotic systems, e.g. graphene sheets and Weyl semi-metals  
\cite{muller-prb-2008, muller-prb-2008b, fritz-prb-2008, muller-prl-2009,
harukazu-jps-2015, hartnoll-arxiv-2016,  lucas-pnas-2016, lucas-prb-2016,
lucas-prb-2016b, crossno-science-2016, bandurin-science-2016, gabbana-cf-2018,
lucas-jop-2018}. 
Graphene is particularly relevant for our analysis, since in this material
charge carriers mimic ultra-relativistic particles \cite{wallace-prl-1947},
positioning itself in a regime of  parameters for which the differences between the 
results of Grad's method of moment and Chapman-Enskog
expansion are larger, as we shall see in the following.

Furthermore, a fascinating connection between hydrodynamics and black hole physics has been established
and intensively explored in the last decade \cite{son-annrev-2007}.
Of particular interest is  the AdS/CFT duality \cite{maldacena-ijtp-1999,aharony-pr-2000}, which 
connects (d+1)-dimensional gravity with d-dimensional field theory \cite{bhattacharyya-jhep-2008}. 
In this framework, fluid dynamic solutions in  (2+1)-dimensions  provide valuable
information for the study of gravity  in (3+1)-dimensions. 
For example, the development of turbulence in (3+1)-dimensional gravitational perturbations 
\cite{adams-prl-2014} has sparked a significant interest for the analysis of relativistic turbulent flows 
in (2+1)-dimensions (\cite{carrasco-prd-2012, green-prx-2014, westernacher-jhep-2017}).

In spite of its importance, a robust methodology connecting kinetic and hydrodynamic 
parameters in (2+1)-dimensions is still lacking; Mendoza et al. \cite{mendoza-jsm-2013} derived transport coefficients for an ultra-relativistic 
ideal gas using Grad's method of moments and the relaxation time approximation (RTA) while, to the best of our knowledge,
the Chapman-Enskog expansion has not been fully derived, with only one calculation of thermal conductivity
available in literature \cite{garcia-perciante-jsp-2017}.

Starting from this state of affairs, in this paper we develop a robust simulation environment 
for viscous relativistic fluid dynamics, based on a two step approach: 
i) a complete theoretical derivation of the transport coefficients of an ideal gas in $(2+1)$ dimensions for 
all kinematic regimes (from ultra-relativistic to near non-relativistic) using both the CE approach and Grad's method; 
ii) a comparison of the predictions of both approaches against accurate numerical simulations, based on a 
recent lattice kinetic scheme \cite{gabbana-pre-2017}. 

Our main results are as follows; i) neat numerical evidence that also in (2+1)-dimensions the CE expansion accurately describes dissipative effects 
in the relativistic regime, while Grad's method fails to do so, and ii) a controlled and systematic procedure relating 
macroscopic  transport parameters to the kinetic relaxation time, thus 
allowing an accurate calibration of the numerical simulations.

Items i) and ii) provide a unified framework  for accurate numerical studies of transport phenomena in relativistic fluids under
quite general conditions, i.e. flows with strong nonlinearities, in non-ideal geometries, across both ultra-relativistic
and near-non relativistic regimes.

This paper is structured as follows: in Section~\ref{sec:hydro} we introduce the relevant equations
describing a relativistic fluid in (2+1)-dimensions at both the mesoscopic and macroscopic levels.
We then sketch the Chapman-Enskog expansion and provide the analytical results of both CE and Grad's method of moments.
In Section~\ref{sec:numerics} we present a numerical analysis giving clear evidence that the transport coefficients
calculated using the Chapman-Enskog expansion provide the correct bridge between the mesoscopic and the macroscopic layers.
To conclude, in Section~\ref{sec:outlook} we summarize our results and future directions of research.

\section{Hydrodynamic derivations}\label{sec:hydro}

In the following, we consider a $(2+1)$ Minkowski space, with metric tensor 
$\eta^{\alpha \beta} = diag(1, -1, -1)$ and use the Einstein summation convention over repeated
indexes, with Latin indexes for 2-D space coordinates and Greek indexes for
$(2+1)$ space-time coordinates. We use natural units, $c = k_B = \hbar = 1$.

Our starting point is the relativistic Boltzmann equation in the RTA
given by the Anderson-Witting model \cite{anderson-witting-ph-1974a,anderson-witting-ph-1974b}:
\begin{equation}\label{eq:rta}
  p^{\alpha} \frac{\partial f}{\partial x^{\alpha}} = \frac{p^{\mu} U_{\mu}}{\tau} \left( f - f^{eq} \right) \quad ;
\end{equation}
the particle distribution function $f( x^{\alpha}, p^{\beta} )$ depends on space-time coordinates 
$x^{\alpha} = \left( t, \bm{x} \right)$ and momenta $p^{\alpha} = \left( p^0, \bm{p} \right) = \left( \sqrt{\bm{p}^2+m^2}, \bm{p} \right)$,
with $\bm{x}, \bm{p} \in \mathbb{R}^2$,
$U^{\alpha}$ is the macroscopic relativistic velocity, $\tau$ is the relaxation (proper-) time, and $f^{\rm eq}$ 
is the equilibrium distribution function, here taken to be the Maxwell-J\"uttner distribution which in $(2+1)$ dimensions writes as
\begin{equation}
  f^{\rm eq} = \frac{n e^{\zeta}}{2 \pi T^2 (\zeta + 1)} e^{- \frac{p^{\mu}U_{\mu}}{\mathstrut{T}} } \quad ;
\end{equation}
$n$ is the particle density, and $\zeta$ is the ratio between the rest mass $m$ and the temperature $T$.
The parameter $\zeta$ physically characterizes the kinematic regime of the macroscopic fluid, with $\zeta \rightarrow 0$ 
in the ultra-relativistic regime and $\zeta \rightarrow \infty$ in the classical one.
The Anderson-Witting model ensures the local conservation of particle number, energy and momentum:
\begin{align}
  \partial_{\alpha} N^{\alpha}              &= 0 \quad , \label{eq:cons1}\\
  \partial_{\beta } T^{\alpha \beta}        &= 0 \quad , \label{eq:cons2}
\end{align}
with $N^{\alpha}$ and $T^{\alpha \beta}$ respectively the particle flow and energy momentum tensors.
These equations are purely formal until a specific form for $N^{\alpha}$ and $T^{\alpha \beta}$ is specified.
The Anderson-Witting model is compatible with the Landau-Lifshitz decomposition \cite{cercignani-book-2002}:
\begin{align}
  N^{\alpha}       = \int f p^{\alpha} \frac{\diff^2 p}{p_0}           &= n U^{\alpha}  - \frac{n}{P + \epsilon} q^{\alpha} \quad , \label{eq:ll-decomposition-o1} \\
  T^{\alpha \beta} = \int f p^{\alpha} p^{\beta} \frac{\diff^2 p}{p_0} &= \epsilon U^{\alpha} U^{\beta} - \left( P + \varpi \right) \Delta^{\alpha \beta} + \pi^{< \alpha \beta >} , \label{eq:ll-decomposition-o2}
\end{align}
$\epsilon$ is the energy density, $P$ the hydrostatic pressure, $q^{\alpha}$ 
is the heat flux, $\pi^{<\alpha \beta>}$ the pressure deviator, $\varpi$ the dynamic pressure,
and $\Delta^{\alpha \beta} = U^{\alpha} U^{\beta} - \eta^{\alpha \beta}$ is the (Minkowski-)orthogonal projector  
to the fluid velocity $U^{\alpha}$; the latter, in the Landau frame, is defined as $T^{\alpha \beta} U_{\beta} = \epsilon U^{\alpha}$.
It is useful to recall that in equilibrium $\varpi = 0$, $q^{\alpha} = 0$ and $\pi^{<\alpha \beta>} = 0$.
On the other hand, the non-equilibrium contribution to the energy momentum tensor can be used to define 
the transport coefficients \cite{cercignani-book-2002}:

\begin{align}
  q^{\alpha}             & =   \lambda  \left( \nabla^{\alpha} T - T U^{\alpha} \partial_{\beta} U^{\beta} \right) \quad , \label{eq:heat-flux} \\
  \pi^{< \alpha \beta >} & =   \eta     \left( \Delta^{\alpha}_{\gamma} \Delta^{\beta}_{\delta} + \Delta^{\alpha}_{\delta} \Delta^{\beta}_{\gamma}  -  \Delta^{\alpha \beta} \Delta_{\gamma \delta} \right) \nabla^{\gamma} U^{\delta} \quad , \label{eq:stress-tensor} \\
  \varpi                 & = - \mu      \nabla_{\alpha} U^{\alpha} \quad ; \label{eq:dynamic-pressure} 
\end{align}
$\lambda$ is the thermal conductivity, $\eta$ and $\mu$ the shear and bulk viscosities, 
and we have used the shorthand notation
\begin{equation}
\begin{array}{lcl} 
  \nabla^{\alpha}         & = & \Delta^{\alpha \beta} \partial_{\beta}       \quad ,\\
  \Delta^{\alpha}_{\beta} & = & \Delta^{\alpha \gamma} \Delta_{\gamma \beta} \quad .
\end{array} 
\end{equation}

The CE expansion allows to define a pathway between kinetic theory and fluid dynamics, 
linking the macroscopic transport coefficients $\lambda$, $\mu$, $\eta$ to the mesoscopic ones, 
in our case the relaxation time $\tau$.

The CE expansion of the relativistic Boltzmann equation was derived several decades ago in $(3+1)$ 
dimensions, see, e.g., \cite{cercignani-book-2002}.
Here we briefly summarize the main steps of the procedure and derive results in
$(2+1)$ dimensions, leaving full mathematical details to an extended version of
this paper.

The starting point is to approximate the one-particle distribution with the sum of two terms,
the equilibrium distribution $f^{\rm eq}$ and a non equilibrium part $f^{\rm neq}$, 
under the assumption that $f^{\rm neq}$ is a small deviation from  equilibrium:
\begin{equation}\label{eq:f-approx}
   f = f^{\rm eq} + f^{\rm neq} = f^{\rm eq} (1 + \phi) \quad ,
\end{equation}
with $\phi$ of the order of the Knudsen number $\rm Kn$, defined as the ratio between the mean free path 
and a typical macroscopic length scale.
From Eq.~\ref{eq:ll-decomposition-o1} and \ref{eq:ll-decomposition-o2} we infer the following constraints 
on the particle distribution function:
\begin{align}
  n &= U^{\alpha} N_{\alpha} = U^{\alpha} \int f^{\rm eq} p_{\alpha} \frac{\diff^2 p}{p_0} \notag \\
                            &= U^{\alpha} \int f          p_{\alpha} \frac{\diff^2 p}{p_0} \quad , \\
  n~e &                      = U^{\alpha} U^{\beta} \int f^{\rm eq} p_{\alpha} p_{\beta} \frac{\diff^2 p}{p_0} \notag \\
                            &= U^{\alpha} U^{\beta} \int f          p_{\alpha} p_{\beta} \frac{\diff^2 p}{p_0} \quad .
\end{align}                            
These conditions together with Eq.~\ref{eq:f-approx} lead to the following constraints for the non-equilibrium part:
\begin{align}
  U_{\alpha}           \int f^{\rm eq} \phi p^{\alpha}            \frac{\diff^2 p}{p_0} = 0 \quad , \label{eq:cond1} \\
  U_{\alpha} U_{\beta} \int f^{\rm eq} \phi p^{\alpha} p^{\beta}  \frac{\diff^2 p}{p_0} = 0 \quad . \label{eq:cond2}
\end{align}
Plugging Eq.~\ref{eq:f-approx} into Eq.~\ref{eq:rta} we obtain
\begin{equation}\label{eq:rta-rbe-ce}
  p^{\alpha} \frac{\partial f^{\rm eq}}{\partial x^{\alpha}} = - \frac{U^{\alpha} p_{\alpha}}{\tau} f^{\rm eq} \phi \quad ,
\end{equation}
where on the LHS we have ignored the term $p^{\alpha} \frac{\partial f^{\rm eq} \phi}{\partial x^{\alpha}}$ since it is $\mathcal{O}(\rm{Kn}^2)$.
We multiply Eq.~\ref{eq:rta-rbe-ce} by $\{1, p^{\beta} \}$, integrate in momentum space and use the result in
combination with Eq.~\ref{eq:cond1} and ~\ref{eq:cond2} to derive the conservation equations:
\begin{equation}
  \begin{aligned}\label{eq:conservation-eqs}
    U_{\alpha} \partial^{\alpha} n + n \nabla^{\alpha} U_{\alpha}             &= 0 \quad , \\
    n c_v U_{\alpha} \partial^{\alpha} T  + P \nabla_{\alpha} U^{\alpha}      &= 0 \quad , \\
    \nabla^{\beta} P  - (P + \epsilon) U_{\alpha} \partial^{\alpha} U^{\beta} &= 0 \quad ,
  \end{aligned}
\end{equation}
where $c_v = (\zeta^2 + 4 \zeta + 2) / (1 + \zeta)^2 $ is the heat capacity at constant volume.
From Eq.~\ref{eq:rta-rbe-ce} we then obtain an expression for $\phi$:
\begin{align}\label{eq:phi-est}
  \phi = -\frac{\tau}{p^\mu U_\mu}p^\alpha 
          \left[  \frac{\partial_\alpha n}{n}
                - \left(1+\zeta+\frac{1}{1+\zeta}\right)\frac{\partial_\alpha T}{T} 
   \notag \right. \\
          \left.
                + p^{\beta}\frac{U_{\beta}\partial_{\alpha}T}{kT^2}
                - \frac{p^{\beta}\partial_{\alpha}U_{\beta}}{kT}
          \right] \quad ;
\end{align}
Next, we apply the projectors $\Delta^{\alpha}_{\beta}$ to $N_{\alpha}$ (Eq.~\ref{eq:ll-decomposition-o1})
and respectively $\Delta_{\alpha \beta}$ and 
$(\Delta_{\beta}^{\gamma} \Delta_{\alpha}^{\delta} - \frac{1}{2} \Delta^{\gamma \delta} \Delta_{\alpha \beta} )$ 
to $T^{\alpha \beta}$ (Eq.~\ref{eq:ll-decomposition-o2}) to obtain:
\begin{align}
  q^{\alpha}           &= -\frac{P + \epsilon}{n} \Delta^{\alpha}_{\beta}N^{\beta} \quad , \label{eq:proj1} \\
  P + \varpi           &= -\frac{1}{2}\Delta_{\alpha\beta}T^{\alpha\beta}          \quad , \label{eq:proj2} \\
  \pi^{<\gamma\delta>} &= \left(\Delta_{\beta}^{\gamma} \Delta_{\alpha}^{\delta} - \frac{1}{2} \Delta^{\gamma \delta} \Delta_{\alpha \beta} \right)T^{\alpha \beta} \quad .\label{eq:proj3}
\end{align}
We now use Eq.~\ref{eq:f-approx} together with Eq.~\ref{eq:phi-est} to calculate $N^{\alpha}$ and $T^{\alpha \beta}$ via 
their integral definitions, eliminate the convective time derivatives using Eqs.~\ref{eq:conservation-eqs},
and obtain the expression of the transport coefficients by direct comparison of Eq.~\ref{eq:proj1},~\ref{eq:proj2} and \ref{eq:proj3} with 
respectively Eq.~\ref{eq:heat-flux}, ~\ref{eq:stress-tensor} and ~\ref{eq:dynamic-pressure}:

\begin{align}
\lambda &=  \tau  n \frac{\zeta ^2+3 \zeta +3}{2 (\zeta +1)^3} \\
 \times & \left( \zeta ^3+ \zeta ^2 \left( 2 -e^{\zeta } \Gamma (0,\zeta ) \left(\zeta ^2+3 \zeta +3\right) \right) +2 \zeta +1 \right) , \notag \\ 
\mu &= \tau P \zeta ^4 \frac{ e^{\zeta } \Gamma (0,\zeta ) \left(\zeta ^2+4 \zeta +2\right) -\zeta -3}{4 \zeta ^3+20 \zeta ^2+24 \zeta +8} ,\\
\eta &= \tau P \frac{e^{\zeta } \Gamma (0,\zeta )\zeta ^4 -\zeta ^3+\zeta ^2+6 \zeta +6 }{ 8 \zeta + 8} ,
\end{align}
where 
\begin{equation*}
  \Gamma(\alpha, x) = \int_x^{\infty} y^{\alpha - 1} e^{-y} \diff y \quad ,
\end{equation*}
is the upper incomplete gamma function.
In the ultra-relativistic limit these expressions simplify to
\begin{align}
\lambda_{\rm ur} &= \frac{3}{2} \tau   n \quad , \\
    \mu_{\rm ur} &= 0                    \quad , \\
   \eta_{\rm ur} &= \frac{3}{4} \tau P   \quad .
\end{align}

For Grad's method, following a procedure similar to those described in \cite{cercignani-book-2002}
for the $(3+1)$ dimensional case, and in \cite{mendoza-jsm-2013} for the
ultra-relativistic $(2+1)$ dimensional case, we obtain the following expressions:
\begin{align}
\lambda &= \frac{\tau  n \left(\zeta ^2+3 \zeta +3\right) \left(2 \zeta ^2+6 \zeta +3\right)^2 }{(\zeta +1)^3 \left(2 \zeta ^4+18 \zeta ^3+57 \zeta ^2+72 \zeta +36\right)} , \\
    \mu &= \tau P \frac{\zeta ^4}{(\zeta +1) \left(\zeta ^2+4 \zeta +2\right) \left(\zeta ^3+9 \zeta ^2+18 \zeta +6\right)} , \\
   \eta &= \tau P \frac{\left(\zeta ^2+3 \zeta +3\right)^2}{(\zeta +1) \left(\zeta ^3+6 \zeta ^2+15 \zeta +15\right)} .
\end{align}
with the ultra-relativistic limit given by:
\begin{align}
\lambda_{\rm ur} &= \frac{3}{4} \tau   n \quad , \\
    \mu_{\rm ur} &= 0                    \quad , \\
   \eta_{\rm ur} &= \frac{3}{5} \tau P   \quad .
\end{align}

These limiting values are the same as those computed by \cite{mendoza-jsm-2013} for $\mu$ and $\eta$, 
while we have a discrepancy of a factor $2$ for $\lambda$.
This discrepancy, whose origin is not clear to us, has no impact on our phenomenological analysis, as we discuss in the following.

%
\begin{figure*}[t]
\centering
\begin{overpic}[width=.98\columnwidth]{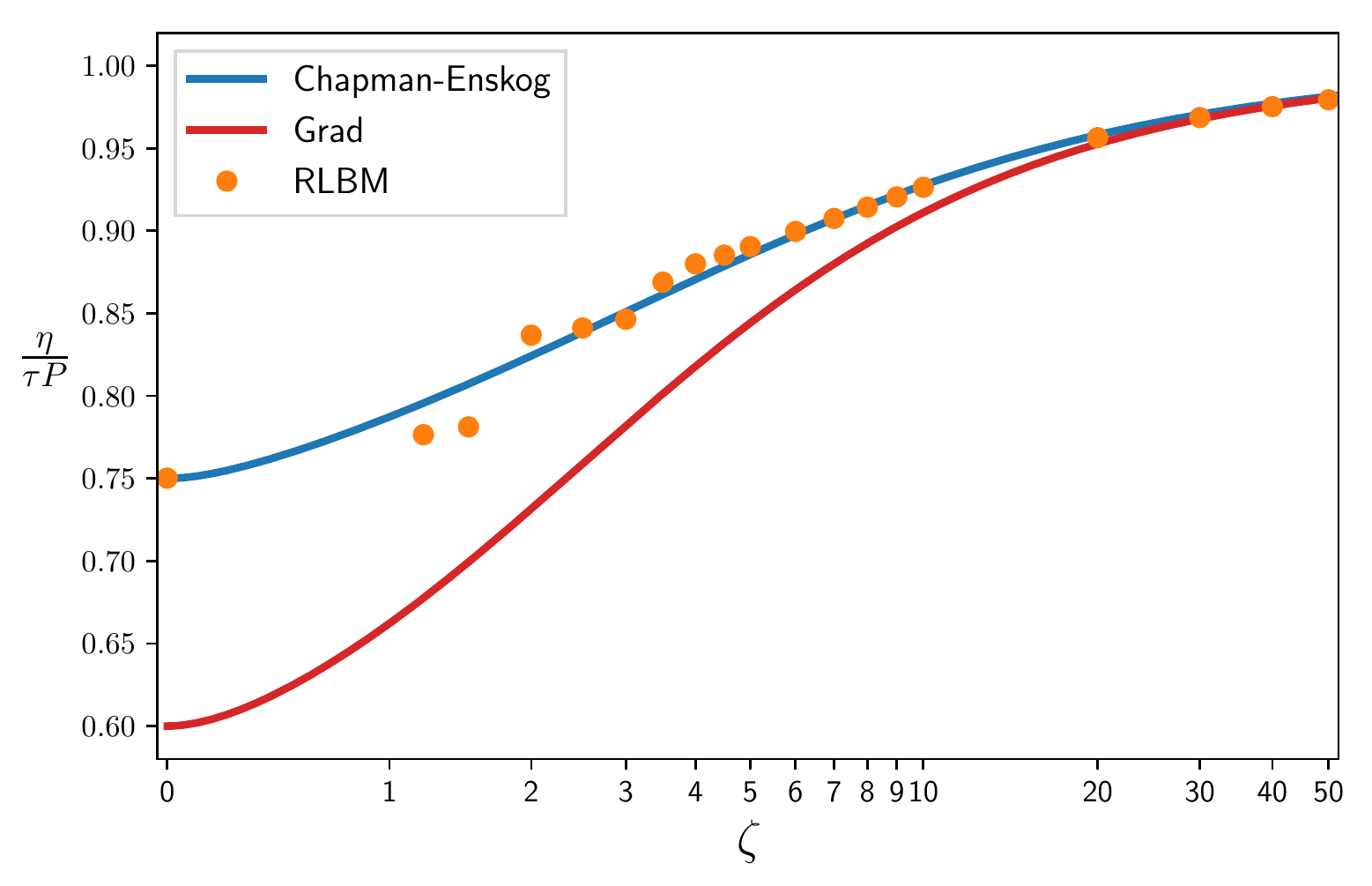}\put(0,58){(a)}\end{overpic}
\begin{overpic}[width=.98\columnwidth]{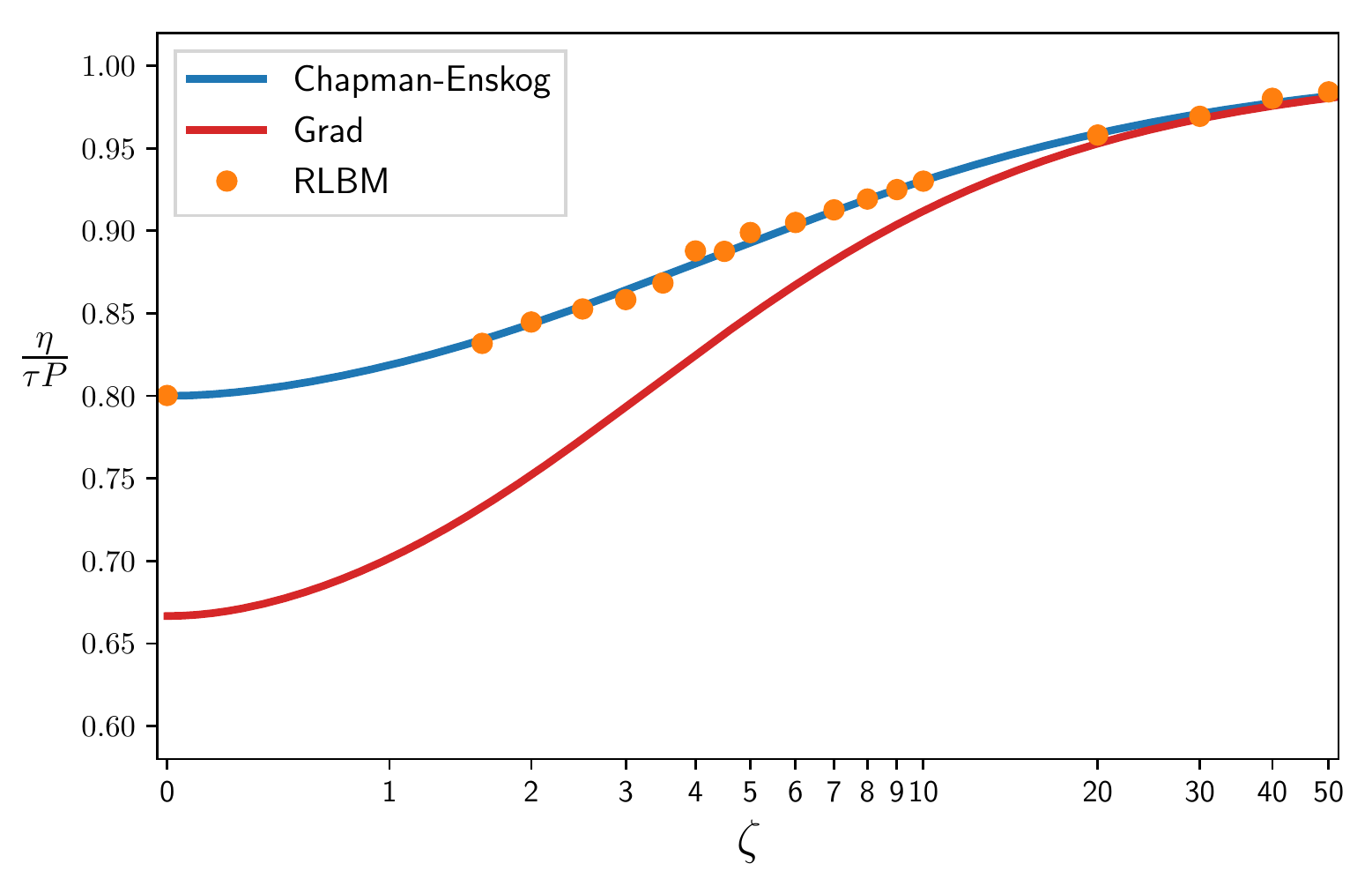}\put(0,58){(b)}\end{overpic}
\begin{overpic}[width=.98\columnwidth]{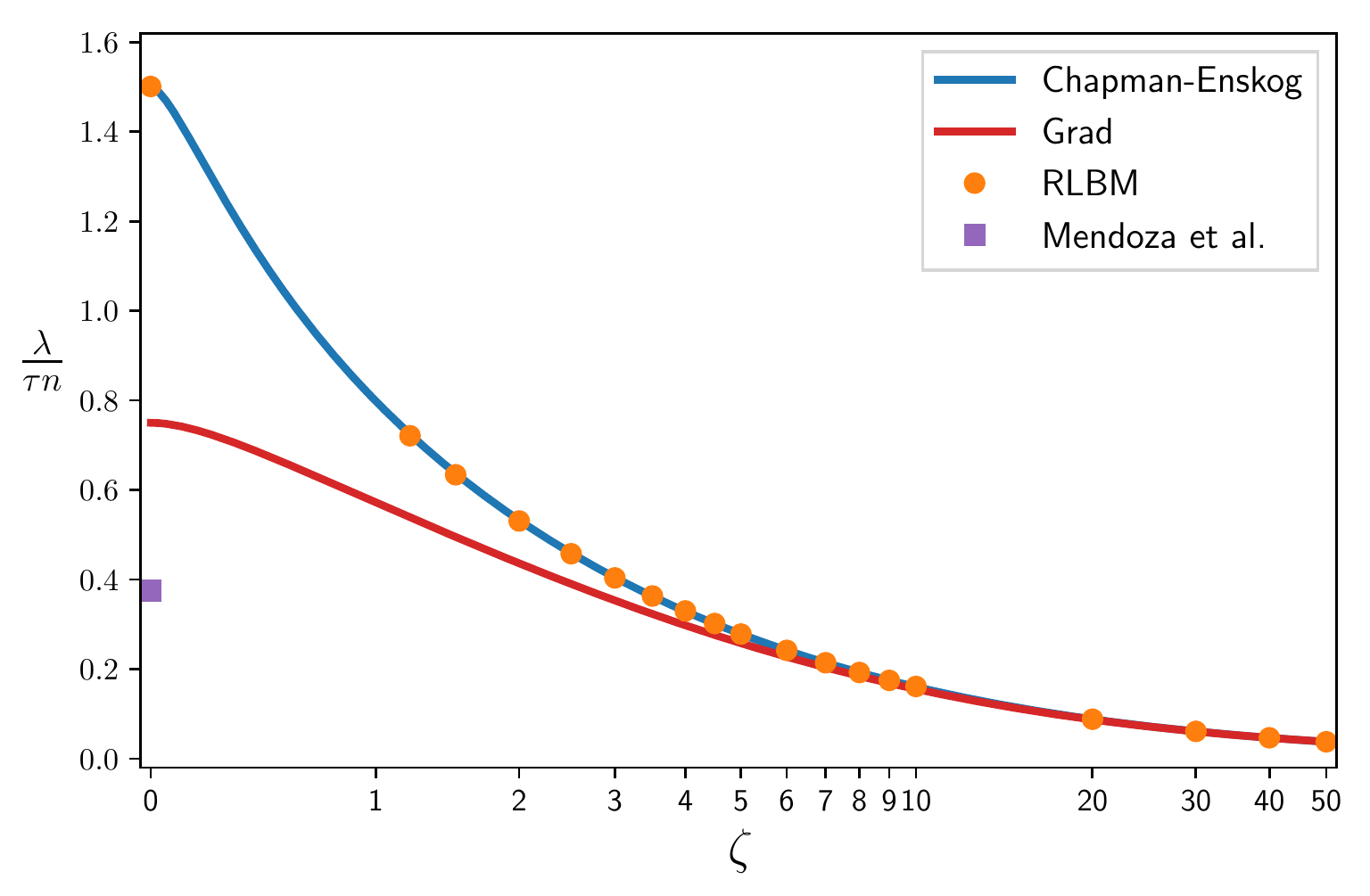}\put(0,58){(c)}\end{overpic}
\begin{overpic}[width=.98\columnwidth]{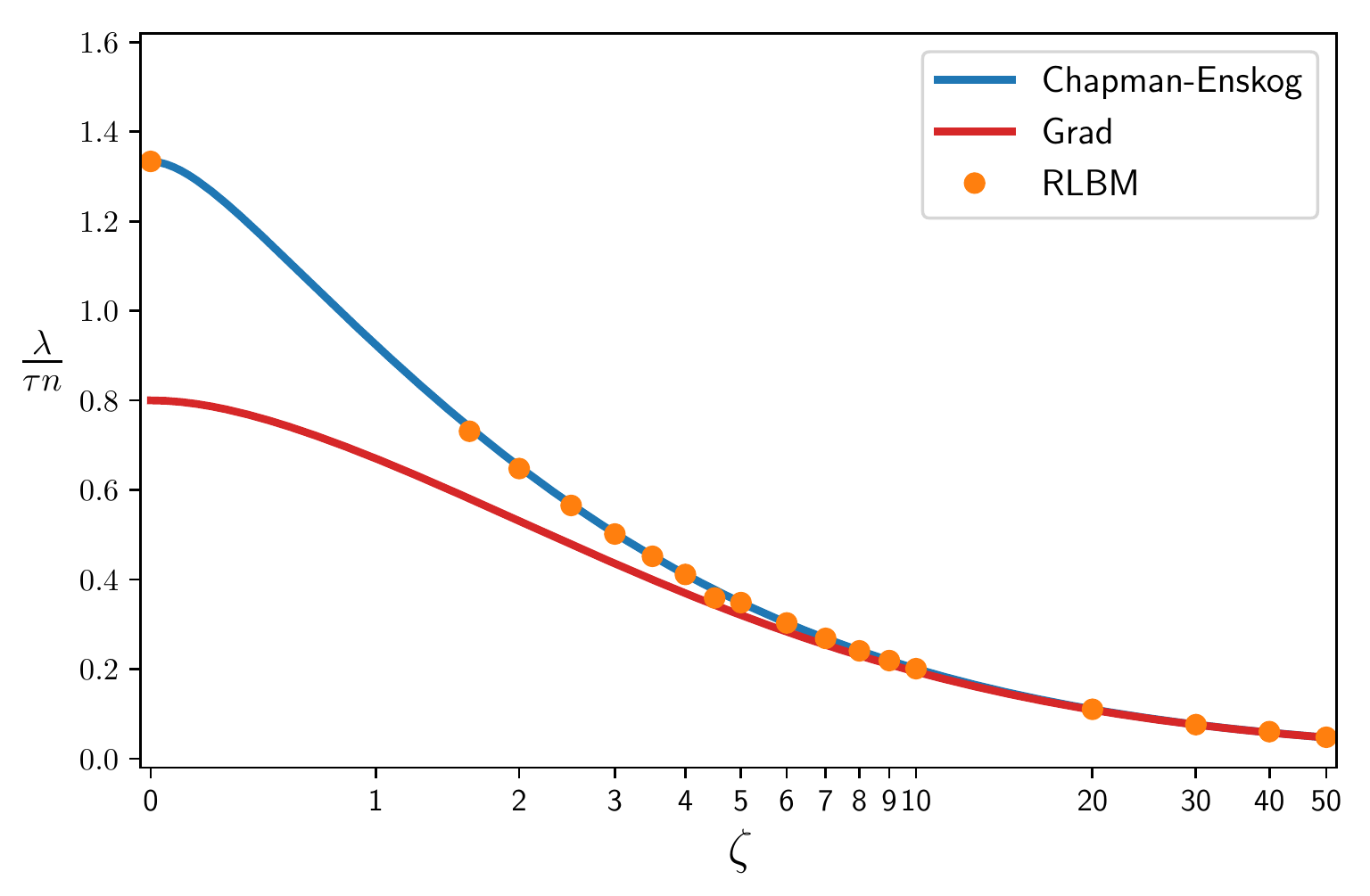}\put(0,58){(d)}\end{overpic}
\begin{overpic}[width=.98\columnwidth]{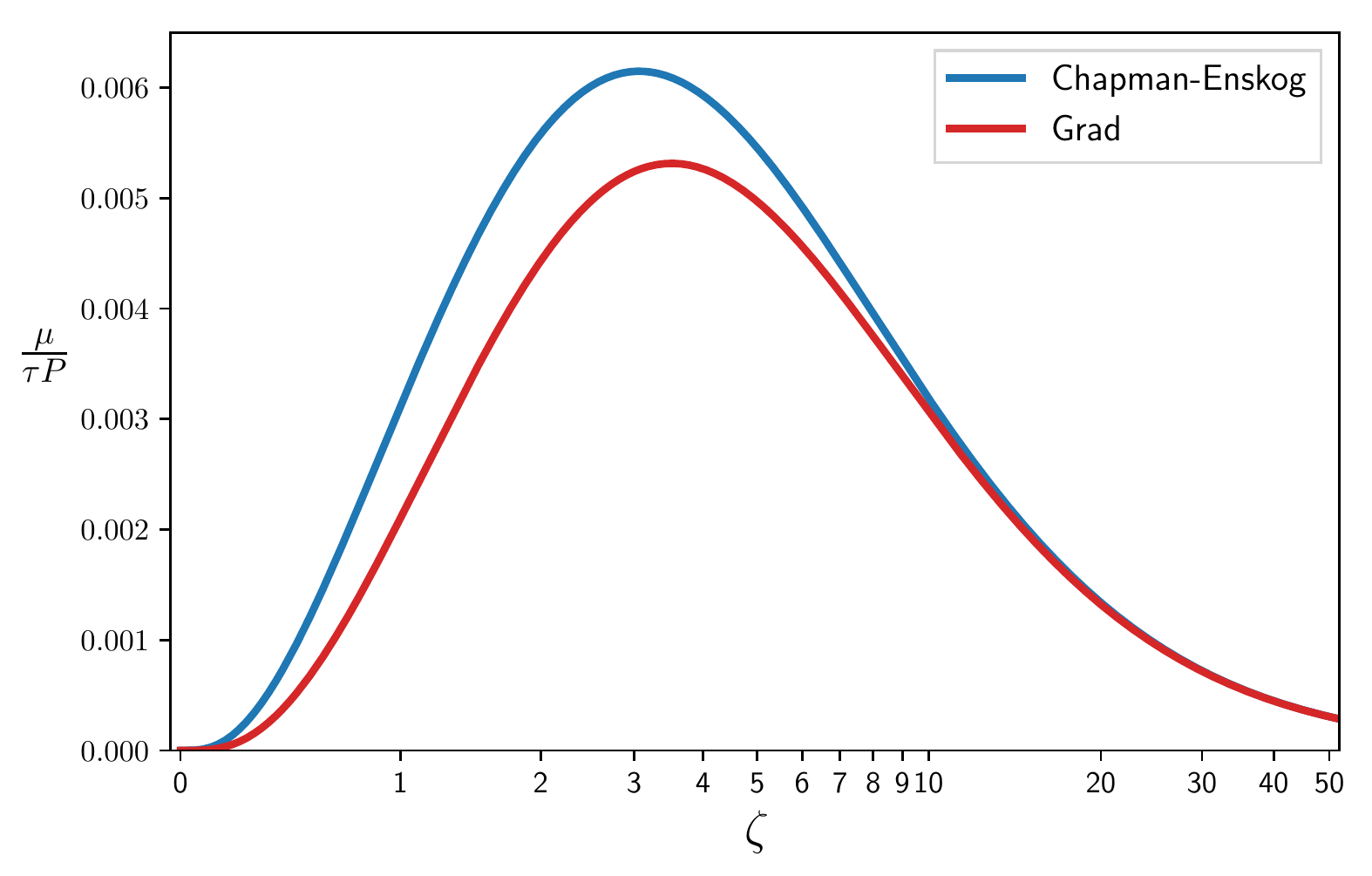}\put(0,58){(e)}\end{overpic}
\begin{overpic}[width=.98\columnwidth]{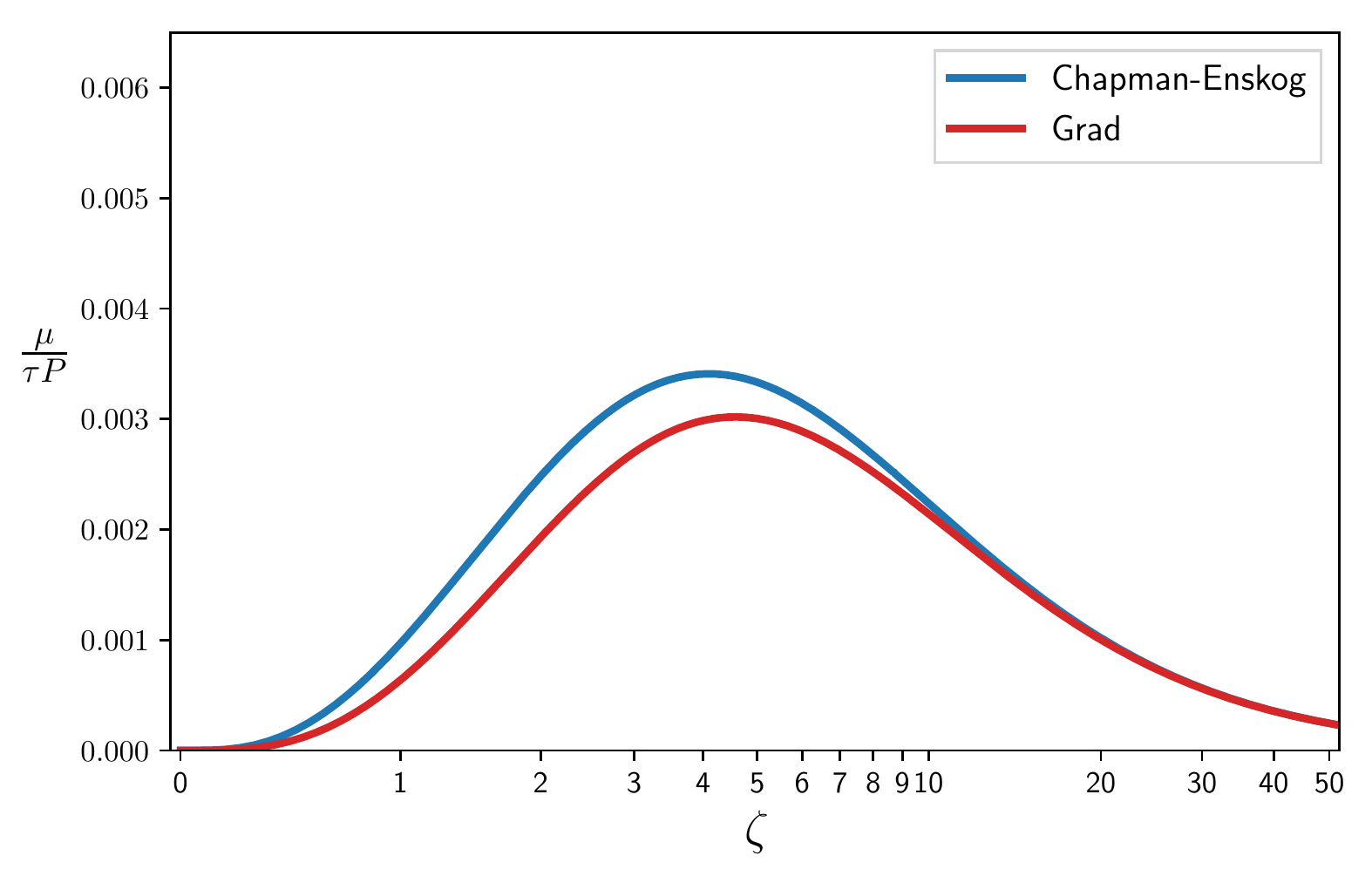}\put(0,58){(f)}\end{overpic}
\caption{ 
    Comparison of the non-dimensional transport coefficients for an ideal
    relativistic gas in $(2+1)$ dimensions (left) and $(3+1)$
    dimensions (right, from \cite{gabbana-pre-2017b}), obtained applying the Chapman-Enskog
    expansion and Grad's method to the relativistic Boltzmann
    equation in the relaxation time approximation. For the thermal
    conductivity $\lambda$ and the shear viscosity $\eta$ we show  the
    results of numerical measurements obtained using a lattice kinetic
    solver \cite{gabbana-pre-2017} which clearly rule in favor of the
    predictions of Chapman-Enskog.  For the bulk viscosity $\mu$ only the
    analytical results are available. We also show (panel (c)) the prediction for
    the ultra-relativistic thermal conductivity in $(2+1)$  dimensions
    by Mendoza et al. in \cite{mendoza-jsm-2013} obtained with
    Grad's method, and differing by a factor two with respect to
    our calculations. Errors are of the order of $1\%$ for all the
    numerical measurements (bars not shown).
        }\label{fig:coeff-cmp}
\end{figure*}
%

\section{Numerical validation}\label{sec:numerics}

Precisely in the same way as in $(3+1)$ dimensions (see \cite{cercignani-book-2002} for details), 
the CE expansion and Grad's method yield different results for the transport coefficients.
In order to discriminate between the two, we perform numerical experiments using a recently developed lattice kinetic scheme \cite{gabbana-pre-2017}.
We consider relativistic flows for which we are able to compute approximate solutions explicitly 
depending on the transport coefficients, and compare with numerical results, obtaining an explicit 
correspondence of the values of the transport coefficients with the relaxation time $\tau$.

First, we consider shear viscosity; we follow \cite{gabbana-pre-2017b} and consider
as a benchmark the Taylor-Green vortex \cite{taylor-prsl-1937}, a well known example of a 
decaying flow with an exact solution of the classic Navier-Stokes equations, and for which 
an approximate solution can be derived in the relativistic regime \cite{gabbana-pre-2017b}.
From the following initial conditions in a 2D periodic domain:
\begin{equation}
  \begin{aligned}\label{eq:tg-classic-initial-conditions}
    u_x(x,y,0) &= \phantom{-}v_0 \cos{\left(x\right)} \sin{\left(y\right)}, \\
    u_y(x,y,0) &=          - v_0 \cos{\left(y\right)} \sin{\left(x\right)}, \quad x,y \in [0, 2 \pi]
  \end{aligned}
\end{equation}
with $v_0$ a initial velocity, it is possible to define the following approximated solution:
\begin{equation}
  \begin{aligned}\label{eq:tg-classic-solution}
    u_x(x,y,t) &= \phantom{-}v_0\cos{\left(x\right)} \sin{\left(y\right)} F(t), \\
    u_y(x,y,t) &=          - v_0\cos{\left(y\right)} \sin{\left(x\right)} F(t), \quad x,y \in [0, 2 \pi]
  \end{aligned}
\end{equation}
with
\begin{equation}
  F(t)  = \exp{ \left( - \frac{2 \eta}{P + \epsilon} t  \right) } \quad ,
\end{equation}
which allows us to numerically measure $\eta$. We perform several simulations
with different value of the relaxation time $\tau$ and fit the coefficient linking
$\eta$ and $\tau$ at different values of $\zeta$.
Fig.~\ref{fig:coeff-cmp}a shows our new results for the non-dimensional shear viscosity in $(2+1)$ dimensions, while  Fig.~\ref{fig:coeff-cmp}b shows results
for the $(3+1)$ dimensional case, previously presented in \cite{gabbana-pre-2017b}.
Our data clearly show that the Chapman-Enskog expansion correctly matches the measured behavior in all regimes, 
while this is not the case for Grad's method.

Further evidence is given when taking into consideration thermal conductivity.
We consider a second benchmark, in which following \cite{coelho-cf-2018},
two parallel plates are kept at constant temperatures, $T_0$ and $T_1$, $T_1 - T_0 = \Delta T$. 
For sufficiently small values of $\Delta T$, and consequently low velocities compared to the speed of light, 
Eq.~\ref{eq:heat-flux} reduces to Fourier's law. 
Under these settings, simulations reach a steady state in which we obtain an approximately constant value 
for the heat flux $q^{\alpha}$, measured via Eq.~\ref{eq:ll-decomposition-o1},
as well as a constant temperature gradient allowing to use Eq.~\ref{eq:heat-flux} to numerically fit $\lambda$.

Results shown in Fig.~\ref{fig:coeff-cmp} are once again in excellent agreement with CE predictions, 
while the results obtained with Grad's are at strong variance with our numerical findings in the mild-relativistic to ultra-relativistic regime.
This conclusion is in no way affected by the discrepancy between our results and those of Mendoza et al. 
\cite{mendoza-jsm-2013} in the ultra-relativistic limit.

Before closing, we wish to spend a few tentative comments on the reasons why relativistic dissipation obeys
Chapman-Enskog asymptotics rather than Grad's expansion.
As mentioned earlier on, the two procedures differ considerably in spirit, before they do in their mathematical
formulation. Grad's expansion is based on a low-order truncated representation of the Boltzmann distribution in Hilbert
space, while the Chapman-Enskog expansion is basically a weak-gradient approximation. 
The recognized weakness of Grad's procedure is that truncation endangers positive-definiteness, while
Chapman-Enskog is, in principle, confined to comparatively mild inhomogeneities, i.e. weak departures
from local equilibrium. 
Other authors have indeed shown \cite{denicol-prd-2012} that extending Grad's method to account for
higher moments, beyond the 14-terms of the standard IS formulation, one eventually 
approaches the CE results.
Since hydrodynamics is a weak-gradient approximation of kinetic theory, 
on purely intuitive grounds, the Chapman-Enskog route appears indeed a 
more natural candidate to describe transport phenomena than Grad's expansion. 
In this respect, it is worth noting that, for all its formal elegance, even for non-relativistic fluids
Grad's has only met with mixed success, while Chapman-Enskog techniques have proved 
significantly more viable (for a detailed discussion see Chapter 6 of \cite{succi-book-2018}).
In other words, even though they provide the same analytical transport coefficients, they 
are {\it not} equivalent at all in practical and numerical terms.
Relativity exposes this gap already at the analytical level.

\section{Conclusions and Future Directions}\label{sec:outlook}

In summary, this paper has presented a complete analytical derivation of the transport coefficients 
of an ideal gas in $(2+1)$ dimensions, encompassing both ultra-relativistic and near 
non-relativistic regimes, for both Chapman-Enskog and Grad's methods. 
A detailed comparison between analytical and numerical results, unambiguously shows 
that relativistic dissipation obeys Chapman-Enskog asymptotics. 
The present works marks a concrete step towards a unified kinetic scheme for 
computational studies of two and three dimensional dissipative relativistic fluid dynamics.
We plan to further extend the present methodology to include quantum statistics, so as 
to perform more detailed studies of hydrodynamic phenomena in graphene \cite{gabbana-prl-2018} and 
other exotic two-dimensional quantum materials \cite{lucas-pnas-2016,moll-science-2016,wang-aip-2017},
including problems related to the AdS/CFT fluid/gravity correspondence \cite{succi-epl-2015}.


\section*{Acknowledgment}

AG has been supported by the European Union's Horizon 2020 research and
innovation programme under the Marie Sklodowska-Curie grant agreement No. 642069.
DS has been supported by the European Union's Horizon 2020 research and
innovation programme under the Marie Sklodowska-Curie grant agreement No. 765048.
SS acknowledges funding from the European Research Council under the European
Union's Horizon 2020 framework programme (No. P/2014-2020)/ERC Grant Agreement No. 739964 (COPMAT).
All numerical work has been performed on the COKA computing cluster at Universit\`a di Ferrara. 


\bibliography{biblio}


\end{document}